\documentclass[a4paper,12pt]{article}
\usepackage{amsmath,amssymb,amsfonts,epsfig,graphicx}
\usepackage{colordvi}

\title{Contribution of the basis-dependent adiabatic 
      geometric phase to noncyclic evolution}

\author{
 M.T. Thomaz\footnote{Corresponding author: mtt@if.uff.br} \\ 
\small\it Instituto de F\'{\i}sica, Universidade Federal Fluminense,\\ 
\small\it Av. Gal. Milton Tavares de Souza s/n$^{\textit o}$, 
CEP 24210-346, Niter\'oi-RJ, Brazil
}

\begin{document}

\maketitle

\begin{abstract}

The geometric phase acquired by the vector
states under an adiabatic evolution along a noncyclic path 
can be calculated correctly in any instantaneous basis of 
a Hamiltonian that varies in time due to a time-dependent 
classical field.
 
\end{abstract}

\vfill

\noindent Keywords: Berry's phase, adiabatic phase, geometric phase,
noncyclic adiabatic evolution, spin-1/2 model.

\noindent PACS numbers: 03.65.Vf, 03.65.Ca

\newpage

%%%%%%%%%%%%%%%%%%%%%%%%%%%%%%%%%%%%%%%%%%%%%%%%%%%%%%%%%%%%
%%%%%%%%%%%%%%%%%%%%%%%%%%%%%%%%%%%%%%%%%%%%%%%%%%%%%%%%%%%%

The important Adiabatic Theorem  demonstrated in 1928 by Born and
Fock\cite{fock} determines the approximate
vector state of quantum systems under adiabatic evolution.
However, only in 1984 Berry\cite{berry} brought
the attention of the Physics community to the adiabatic phases
in vector states driven by  Hamiltonians that 
depend on adiabatically evolving external periodic classical fields.
Berry showed that for
time intervals equal to complete periods of the 
Hamiltonian, a vector state which at $t=0$
is one of the eigenstates of the Hamiltonian acquires
a geometric phase (known as {\it Berry's phase})
that is independent of the complete instantaneous basis 
of the Hamiltonian. Such independence of the adiabatic 
phase is related to the time evolution of the 
state in a time interval that corresponds to a closed
path in the space of the external classical field\cite{berry}.
Samuel and Bhandari\cite{samuel} generalized the 
geometric phase to noncyclic evolution in  1988 in order 
that this phase could also be basis-independent. A great
deal has been done in order to define a geometric
phase in noncyclic adiabatic evolution that is 
basis-independent\cite{gonzalo,zhuprl,zhuprB}.
The existence of these noncyclic geometric phase has been 
verified experimentally\cite{filipp}.

In Ref.\cite{mtthomaz} it was shown that no special recipe  
was needed to calculate a noncyclic adiabatic phase 
to obtain the correct result of measurable physical quantities
in a quantum state evolving under the action of any 
non-degenerate adiabatic Hamiltonian.

In this communication we present the phase acquired by any vector state
describing a quantum system driven by a Hamiltonian that depends 
on a classical set of external parameters
(a classical field) that vary
adiabatically in time. We assume that during the interval of time that
the quantum system evolves the set of classical parameters does not
return to the initial set of values. We also suppose that the energy
spectrum of the Hamiltonian is non-degenerate at any instant $t$.

Let ${\bf H}(\vec{\bf R}(t))$ be the adiabatic Hamiltonian 
whose variation in time comes from the presence of the external 
classical  field $\vec{\bf R} (t)$ with $m$ components 
($\vec{\bf R} (t) \equiv (X_1, X_2, \cdots , X_m)$). 
The field $\vec{\bf R} (t)$ evolves adiabatically.

Let $\{|\varphi_n; t \rangle, n= 1,2, \cdots\}$ be a complete
instantaneous orthonormal basis of ${\bf H}(\vec{\bf R}(t))$, that 
is,

\begin{eqnarray} \label{1}  
{\bf H}(\vec{\bf R}(t)) \; |\varphi_n; t \rangle = E_n (t)\; |\varphi_n; t \rangle 
     \hspace{0.5cm} \mbox{and} \hspace{0.5cm}
\langle \varphi_n; t | \varphi_m; t \rangle = \delta_{n m},
\end{eqnarray}

\noindent where  $n, m = 1, 2, \cdots$. 

Let the initial state of the quantum system be described by the 
vector state $|\psi (0) \rangle$, with 
$\langle \psi (0)| \psi (0) \rangle =1$. This state is independent
of any complete basis in which we decide to project it, and its  
time evolution is given by the Schr\"odinger equation,

\begin{eqnarray}    \label{2}
{\bf H}(\vec{\bf R}(t)) |\psi (t) \rangle = i \hbar 
   \frac{\partial |\psi (t) \rangle}{\partial t}.
\end{eqnarray}

It is simple to solve eq.(\ref{2}) with the initial state
$|\psi (0)\rangle$, in the adiabatic regime, if
we decompose $|\psi (0)\rangle$ in an instantaneous
basis of ${\bf H}(\vec{\bf R}(0))$, 

\begin{eqnarray}  \label{3}
|\psi (0) \rangle = \sum_{j=1}^{M}  a_j \, |\varphi_j; 0 \rangle,
\end{eqnarray}

\noindent with $M>1$, $a_j \in {\mathbb C}$, $j = 1, 2, \cdots M$,
and $\sum_{j=1}^{M} |a_j|^2 = 1$. To  obtain the vector
state $|\psi(t)\rangle$, at any time $t>0$, we apply the Adiabatic 
Theorem\cite{fock,messiah}  

\begin{eqnarray}  \label{4}
|\psi (t) \rangle \approx \sum_{j=1}^{M} a_j \, e^{i \gamma_j (t)}
   \, e^{-\frac{it}{\hbar} \langle E_j (t) \rangle }  \;
         |\varphi_j; t \rangle,
\end{eqnarray}   

\noindent  in which $\gamma_j (t)$ is the adiabatic phase, 

\begin{subequations}

\begin{eqnarray}   \label{5a}
   \gamma_j (t) = i \int_0^t  dt^{\prime} \, \langle \varphi_j; t^{\prime}| 
\left(\frac{d}{dt^{\prime}}| \varphi_j;t^{\prime} \rangle \right),
\end{eqnarray} 

\noindent and $\langle E_j (t) \rangle$ is the average energy of 
the quantum  system from the initial moment ($t=0$)  up to $t$,

\begin{eqnarray} \label{5b}
\langle E_j (t) \rangle \equiv \frac{1}{t} \, \int_0^t dt^{\prime} \, E_j (t^{\prime}),
\end{eqnarray}

\end{subequations}

\noindent  where $j = 1, 2, \cdots, M$.

For an Hamiltonian whose time-dependence comes through the fiel $\vec{R} (t)$, 
the adiabatic phase is rewriten  as\cite{berry} a path integral in 
$\vec{\bf R}$-space,

\begin{eqnarray}   \label{6}
\gamma_j (t) = i \int_{\vec{\bf R}(0)}^{\vec{\bf R}(t)}  \; d\vec{\bf R} \cdot\; 
  \langle \varphi_j; \vec{\bf R}| \left( \vec{\nabla}_{\vec{\bf R}} \; 
  | \varphi_j; \vec{\bf R} \rangle \right)
    \hspace{0.5cm}      j = 1, 2, \cdots, M. 
\end{eqnarray}

\noindent  Here we use the notation: 
$\vec{\nabla}_{\vec{\bf R}} \bullet \equiv
 \frac{\partial \bullet} {\partial X_1} \hat{\imath}_1 + \cdots
 +  \frac{\partial \bullet} {\partial X_m} \hat{\imath}_m$, where
$\hat{\imath}_l$, $ l =1, 2, \cdots, m$, is the unitary vector
along the axis-$X_l$, $l= 1, 2, \cdots, m$, in $\vec{\bf R}$-space.
The relation between the time integral on the r.h.s. of eq.(\ref{5a}) 
and the path integral on the r.h.s. of eq.(\ref{6}) is unique.

The states $| \psi(0) \rangle$ and $|\psi (t) \rangle$ are independent
of the particular choice of instantaneous basis for the 
Hamiltonian; however, once we use the decomposition (\ref{3}) 
to write those states, the coefficients $a_j$, 
$j= 1, 2, \cdots, M$,  depend on the particular instantaneous
basis of energy used.

For a quantum state $|\psi (t)\rangle$ given by eq. (\ref{4}),
the expectation value of any operator ${\bf O}$ (time-dependent 
or not) associated  to a physical quantity has interference effects
due to the presence of  phases\cite{mtthomaz}

\begin{eqnarray}   \label{7}
a_j \, a_k^{*} 
  e^{ i [\gamma_j (t) - \gamma_k (t)]} \;\; 
    e^{- \frac{it}{\hbar} \left[\langle E_j(t) \rangle -  \langle E_k(t) \rangle\right]} \;\;
       \langle \varphi_k; t| {\bf O}| \varphi_j; t\rangle,
\end{eqnarray}

\noindent in which $j, k= 1, 2, \cdots,M$. We should note  that the 
differences of the adiabatic phases, $\gamma_j (t) - \gamma_k (t)$, are
obtained from integrals along the same path in the classical external
parameters space, but followed in the  opposite sense. The sum 
of these two path integrals to calculate  the difference 
$(\gamma_j (t) - \gamma_k (t))$ is not equal to the path 
integral along a closed loop in $\vec{\bf R}$-space.

If instead of using the instantaneous basis 
$\{ |\varphi; t \rangle, n= 1, 2, \cdots\}$,
we had chosen another instantaneous  orthonormal basis 
of ${\bf H} (\vec{\bf R} (t))$, that is,
$\{ |\Phi_n; t \rangle, n= 1, 2, \cdots\}$, where

\begin{eqnarray}  \label{8}
 | \Phi_n; t\rangle = e^{ i \alpha_n (t)} \; | \varphi_n; t \rangle ,
      \hspace{1cm} n = 1, 2, \cdots
\end{eqnarray}

\noindent and the function $\alpha_n (t) \in {\mathbb R}$, 
$n = 1, 2, \cdots$, the initial vector state $|\psi (0) \rangle$
would have the following decomposition

\begin{subequations}

\begin{eqnarray} \label{9a}
|\psi (0)\rangle = \sum_{j=1}^{M} \tilde{a}_j | \Phi_j; 0\rangle,
\end{eqnarray}

\noindent where

\begin{eqnarray}  \label{9b}
 \tilde{a}_j =  e^{-i \alpha_j(0)} \; a_j.
\end{eqnarray}

\end{subequations}

In the following we discuss the term (\ref{7}) in the new instantaneous
basis $\{ |\Phi_n; t \rangle;  n= 1, 2, \cdots \}$:

\vspace{0.2cm}

\noindent 1) the difference of the dynamical phases,
$ e^{- \frac{it}{\hbar} \left[\langle E_j(t) \rangle -  \langle E_k(t) \rangle\right]}$,
for $k, j \in \{1, 2, \cdots, M \}$, is independent of
the instantaneous basis of ${\bf H} (\vec{\bf R} (t))$ applied to 
write the initial state $|\psi (0)\rangle$.

\vspace{0.2cm}

\noindent 2) the product 
$ a_j \, a_k^{*} \; \langle \varphi_k; t| {\bf O}| \varphi_j; t\rangle$ 
becomes:

\begin{eqnarray}    \label{10}
\tilde{a}_j \, \tilde{a}_k^{*} 
       \langle \Phi_k; t| {\bf O}| \Phi_j; t\rangle =   
a_j \, a_k^{*} 
  e^{ -i [\alpha_j (0) - \alpha_k (0)]} \;\; 
     e^{ i [\alpha_j (t) - \alpha_k (t)]} \;\;
       \langle \varphi_k; t| {\bf O}| \varphi_j; t\rangle,
\end{eqnarray}

\noindent showing that this part of the term (\ref{7}) is 
basis-dependent.

\vspace{0.2cm}

\noindent 3) Let $\tilde{\gamma}_j (t)$, $j= 1,2, \cdots, M$, be the 
geometric phases in the new instantaneous basis of energy

\begin{eqnarray}   \label{11}
\tilde{\gamma}_j (t) &=& i \int_{\vec{\bf R}(0)}^{\vec{\bf R}(t)}  \; d\vec{\bf R} \cdot\; 
  \langle \Phi_j; \vec{\bf R}| \left( \vec{\nabla}_{\vec{\bf R}} \; 
  | \Phi_j; \vec{\bf R} \rangle \right)     \nonumber   \\
&=& \gamma_j (t) - \int_{\alpha_j (0)}^{\alpha_j (t)}  \; d\alpha_j.
\end{eqnarray}

\noindent  For a noncyclic evolution, $\vec{\bf R}(t) \not= \vec{\bf R}(0)$,
the path integral in the previous relation is not a loop 
in $\vec{\bf R}$-space.

The relation between the geometric phases  $\gamma_j$ and
$\tilde{\gamma}_j (t)$, $j= 1,2, \cdots, M$, is given by

\begin{eqnarray}    \label{12}
\tilde{\gamma}_j (t) - \tilde{\gamma}_k (t)
=  \gamma_j (t) - \gamma_k (t) - (\alpha_j (t) - \alpha_j (0))
+ (\alpha_k (t) - \alpha_k (0)),
\end{eqnarray}

\noindent showing that this difference depends on the chosen 
instantaneous basis ${\bf H} (\vec{\bf R})$ to decompose 
the initial vector state $|\psi (0) \rangle$. We are correcting our 
statement in Ref.\cite{mtthomaz} where we affirmed that this difference
of adiabatic phases ``is gauge invariant at any time $t$".
However the dependence of $\tilde{\gamma}_j (t) - \tilde{\gamma}_k (t)$
on the functions $\alpha_j (t)$ and $\alpha_k (t)$ cancels 
out the contributions of these functions  in eq.(\ref{10}), 
showing that the result of the term (\ref{7}) does not depend on
the instantaneous  basis of the Hamiltonian. 

It is very simple to show that the vector state $|\psi (t) \rangle$,
at any time $t$, is  also independent of the instantaneous basis
of the Hamiltonian, as it should be.

No special recipe is needed, then, to
 calculate  the adiabatic 
geometric phases $\gamma_j (t)$, $ j= 1, 2, \cdots, M$,
in the state $| \psi (t) \rangle$  along any noncyclic 
adiabatic evolution. Those phases are fundamental
to the determination of the correct quantum state along the 
adiabatic evolution of the quantum system.

\vspace{0.5cm}

We revisit the simple two-level model\cite{ajp2000} 
to exemplify the previous result. The Hamiltonian of the 
spin-$1/2$ in the presence of an external classical
magnetic field $\vec{B} (t)$ is \cite{mtthomaz},
 
\begin{eqnarray}   \label{13}
{\bf H} (t) = \frac{\mu \hbar}{2} \vec{B} (t)\cdot \vec{\sigma},
\end{eqnarray}

\noindent  in which $\vec{B} (t) = (B_x (t), B_y (t), B_z(t))$ is 
the classical time-dependent  magnetic field,  $\mu= g \mu_B$, 
$\mu_B$ is the Bohr magneton and $g$ is the 
Land\'e's factor. The $\vec{\sigma} = ( \sigma_x, \sigma_y, \sigma_z)$,
in which $\sigma_i$, $i \in \{x, y, z\}$  are the Pauli's matrices.
The eigenvalues of Hamiltonian (\ref{13}) are
 $E_1 = - \frac{\mu \hbar B(t)}{2}$
and $E_2 = \frac{\mu \hbar B(t)}{2}$, with $B(t) \equiv |\vec{B} (t)|$.
The respective normalized instantaneous eigenvectors are:

\begin{subequations}

\begin{eqnarray} 
|\varphi_1; t\rangle_f &=& e^{i f(t)} \sqrt{\frac{B(t) - B_z(t)}{2 B(t)}}
 \left[  |\uparrow\rangle -  \frac{(B_x(t) +i  B_y (t))}{B(t) - B_z (t)} |\downarrow\rangle \right],
                  \label{14a} \\
\nonumber \\
%
%segunda linha
%
|\varphi_2; t\rangle_g &=&  e^{i g(t)} \sqrt{\frac{B(t) + B_z(t)}{2 B(t)}}
 \left[  |\uparrow\rangle + \frac{(B_x(t) + i B_y (t))}{B(t) + B_z (t)} |\downarrow\rangle \right].
                  \label{14b} 
\end{eqnarray}

\end{subequations}

\noindent We are using the notation $|\uparrow\rangle$ and $|\downarrow\rangle$
for the eigenvectors of $\sigma_z$, that is, 
$\sigma_z |\uparrow\rangle = + |\uparrow\rangle$  and
$\sigma_z |\downarrow\rangle = - |\downarrow\rangle$. The functions
$f(t)$ and $g(t) \in \mathbb{R}$  are arbitrary continuous
functions.

Consider in what follows the particular type of classical magnetic 
field $\vec{B}(t)$, namely

\begin{eqnarray}  \label{15}
\vec{B} (t) = ( B \sin(\theta) \cos(\omega t), 
	  B\sin(\theta) \sin(\omega t), B\cos(\theta)). 
\end{eqnarray}
	
\noindent with constant modulus $B$, an angle $\theta$ with 
respect to the $z$-axis, and precessing around the latter 
with constant angular frequency $\omega$.
We calculate the geometric phases in the adiabatic evolution
of the vector state $|\psi (t) \rangle$ for the initial vector
$|\psi (0) \rangle$, 

\begin{eqnarray} \label{16}
|\psi (0)\rangle = a_1 | \varphi_1; 0\rangle_f 
      + a_2 |\varphi_2; 0\rangle_g,
\end{eqnarray}

\noindent in which  $a_1$ and $a_2 \in \mathbb{C}$, and
$|a_1|^2 + |a_2|^2 =1$, and  $f(t)$ and $g(t)$ are arbitrary
continuous real functions.

At any instant $t$, the vector state $|\psi (t) \rangle$ is

\begin{eqnarray} \label{17}
|\psi (t) \rangle \approx  a_1 \; e^{i\gamma_1^{(f)} (t)} \; e^{i \frac{\mu B t}{2}} \; 
|\varphi_1; t\rangle_f
 + a_2 \; e^{i\gamma_2^{(g)}  (t)} \; e^{-i \frac{\mu B t}{2}} 
    \;\; |\varphi_2; t\rangle_g.
\end{eqnarray}

The geometric phases calculated in the instantaneous basis
(\ref{14a}) and (\ref{14b}) are 

\begin{subequations}

\begin{eqnarray}
 \gamma_1^{(f)} (t) &=& - \int_{f(0)}^{f(t)} \, df 
     - cos^2 \left(\frac{\theta}{2}\right) \times \omega t,   \label{18a}   \\
\gamma_2^{(g)} (t) &=& - \int_{g(0)}^{g(t)} \, dg 
     - sin^2 \left(\frac{\theta}{2}\right) \times \omega t.   \label{18b}
\end{eqnarray}

\end{subequations}

In Ref.\cite{mtthomaz} the eigenstates of Hamiltonian (\ref{13}) 
correspond to the choices: $f(t) = \pi$ and $g(t) =1$. For these 
functions, eqs.(\ref{17}), (\ref{18a}) and (\ref{18b})  recover
eq.(15) of Ref.\cite{mtthomaz} of the vector state of the 
adiabatic evolution of vector states driven by the Hamiltonian 
(\ref{13}).

\vspace{0.5cm}

In summary, for a noncyclic adiabatic evolution of vector states 
under the action of a Hamiltonian with time-dependent
classical fields, we can use any instantaneous basis of the Hamiltonian 
to calculate the geometric phases once the initial vector state
is decomposed in the same basis as we calculate the geometric phases. 
The differences between those geometric phases are not
basis-independent as have been long pursued in the 
literature\cite{samuel,gonzalo}. Those geometric phases 
have to be taken into account in the correct evaluation of the
contribution to the adiabatic evolution of 
the vector state and the measured quantities in these quantum states.
The presence of the phases and their interference effects are
important in quantum computation.

%%%%%%%%%%%%%%%%%%%%%%%%%%%%%%%%%%%%%%%%%%%%%%%%%%%%%%%%%%%%%%%%%
%%%%%%%%%%%%%%%%%%%%%%%%%%%%%%%%%%%%%%%%%%%%%%%%%%%%%%%%%%%%%%%%%
\vspace{1cm}

The author thanks Eduardo V. Corr\^ea Silva for the careful reading
of this communication.

%%%%%%%%%%%%%%%%%%%%%%%%%%%%%%%%%%%%%%%%%%%%%%%%%%%%%%%%%%%%%%%%
%%%%%%%%%%%%%%%%%%%%%%%%%%%%%%%%%%%%%%%%%%%%%%%%%%%%%%%%%%%%%%%%

\end{document}